\documentclass[sigconf,screen]{acmart}

\AtBeginDocument{%
  \providecommand\BibTeX{{%
    \normalfont B\kern-0.5em{\scshape i\kern-0.25em b}\kern-0.8em\TeX}}}

\setcopyright{rightsretained}
\acmConference[SIGIR eCom'21]{ACM SIGIR Workshop on eCommerce}{July 15, 2021}{Virtual Event, Montreal, Canada}
\acmYear{2021}
\copyrightyear{2021}
\makeatletter
\renewcommand\@formatdoi[1]{\ignorespaces}
\makeatother
\acmISBN{}

\acmPrice{}
\acmDOI{}

\usepackage[belowskip=-10pt,aboveskip=2pt]{caption}

\usepackage[ruled,vlined]{algorithm2e}

\usepackage{cleveref}

\crefformat{section}{\S#2#1#3} 
\crefformat{subsection}{\S#2#1#3}
\crefformat{subsubsection}{\S#2#1#3}

\begin{document}

\title{Transformers with multi-modal features and post-fusion context for e-commerce session-based recommendation}


\author{Gabriel de Souza P. Moreira}
\email{gmoreira@nvidia.com}
\affiliation{
  \institution{NVIDIA}
  \state{S\~ao Paulo}
  \country{Brazil}
}

\author{Sara Rabhi}
\email{srabhi@nvidia.com}
\affiliation{
  \institution{NVIDIA}
  \state{Ontario}
  \country{Canada}
}

\author{Ronay Ak}
\email{ronaya@nvidia.com}
\affiliation{
  \institution{NVIDIA}
  \state{Florida}
  \country{United States}
}

\author{Md Yasin Kabir}
\email{mkabir@nvidia.com}
\affiliation{
  \institution{NVIDIA}
  \state{Missouri}
  \country{United States}
}

\author{Even Oldridge}
\email{eoldridge@nvidia.com}
\affiliation{
  \institution{NVIDIA}
  \state{British Columbia}
  \country{Canada}
}

\renewcommand{\shortauthors}{}

\begin{abstract}
Session-based recommendation is an important task for e-commerce services, where a large number of users browse anonymously or may have very distinct interests for different sessions. In this paper we present one of the winning solutions for the Recommendation task of the SIGIR 2021 Workshop on E-commerce Data Challenge. Our solution was inspired by NLP techniques and consists of an ensemble of two Transformer architectures -- Transformer-XL and XLNet -- trained with autoregressive and autoencoding approaches. To leverage most of the rich dataset made available for the competition, we describe how we prepared multi-model features by combining tabular events with textual and image vectors. We also present a model prediction analysis  to better understand the effectiveness of our architectures for the session-based recommendation.
\end{abstract}

\maketitle

\section{Introduction}

In many recommendation domains such as e-commerce, news, streaming video and music services, users might be untrackable, their histories can be short, and users can have rapidly changing tastes\cite{tagliabue2020shopping}. Providing recommendations based purely on the interactions that happen in the current session is an extremely important and challenging problem. Many methods have been proposed that leverage the sequence of interactions that occur during a session, including session-based k-NN (k-Nearest Neighbours) algorithms like V-SkNN \cite{ludewig2018evaluation} and neural approaches like GRU4Rec. \cite{hidasi2018recurrent}. 

Within the NLP domain, approaches based on Transformers architectures \cite{vaswani2017attention} have demonstrated significant advantages over sequential and CNN based approaches.  Transformers apply self-attention to the input sequence, allowing the model to focus on the most important inputs, while also providing improved access to inputs that occur further back in the sequence. These improvements are important when comparing the effectiveness of Transformers relative to RNN-based models that have limited access to the history of the sequence. 

In this paper we present our solution as one of the winners for the Recommendation task of the SIGIR 2021 Workshop on E-commerce Data Challenge \cite{tagliabue2021sigir}. It consists of an ensemble of two different Transformer architectures -- Transformer-XL and XLNet -- trained with autoregressive and autoencoding approaches inspired by Natural Language Processing (NLP). We leveraged the rich information provided by the dataset, and explored different ways to combine tabular data of user interactions events (e.g., clicks, add-to-card, remove-from-cart, purchases, search queries) with unstructured data (product description and images) in a multi-modal approach.

In \cref{sec:related_work} we introduce the related work, in \cref{sec:challenge} describe the competition and dataset, and in subsequent sections we describe our solution for the competition, including data preprocessing and feature engineering  in \cref{sec:preproc_feat_eng}, model architectures, training and evaluation in \cref{sec:model_train_eval}, and finally our ensembling approach and results in \cref{sec:ensembling}.


\section{Related work}
\label{sec:related_work}
In this section, we describe previous work on sequential and session-based recommendation tasks and also how Transformers architectures, originally proposed for NLP, have been adapted for recommendation tasks.

\vspace{-1.4mm}
\subsection{Sequential and Session-based Recommendation}


Sequential recommendation approaches have been proposed to capture sequential and temporal dependencies between user interactions, including algorithms like Markov models \cite{he2009web,garcin2013personalized,kapoor2015just} and neural architectures like Recurrent Neural Networks (RNNs) \cite{hidasi2015session}, Convolutional Neural Networks (CNNs) \cite{tang2018personalized, wang2019towards} and attention-based networks \cite{wang2020time}. More recently Transformer architectures \cite{wu2020sse,lin2020fissa,sun2019bert4rec,kang2018self,zhang2019next,chen2020improving} have been applied for CTR prediction and next-item recommendation leveraging the sequence of past user interactions. In many real-world applications, however, such longer-term information is often not available, because users are not logged in or because they are first-time users. To address the issue, session-based recommender systems \cite{malte2020empirical} have been proposed to model the sequence of interactions within the current user session, with no access to past user interactions. In this context, a session is a short sequence of user interactions, typically bounded by user inactivity. 

Algorithms based on Session k-NN \cite{malte2020empirical} and neural networks like RNNs \cite{hidasi2015session,moreira2018news,gabriel2019contextual}, CNNs \cite{yuan2019simple}, and memory networks \cite{mi2020memory} have been explored for session-based recommendation. 

\subsection{Transformers for Session-based recommendation}

More recently, Transformers and the self-attention mechanism were also explored for session-based recommendation \cite{chen2019bert4sessrec,zhang2020preference,xu2019graph,pan2020rethinking,sun2019self,luocollaborative,fang2021session}. Most of those works use a causal Language Model (LM) approach for training (auto-regressive), i.e., using only interactions before the target item as input. Only \cite{chen2019bert4sessrec} use a training scheme similar to BERT \cite{devlin2018bert, bianchi2020bert} -- masked LM (autoencoding) -- which randomly masks items in the sequence for prediction, allowing during training the usage of future interactions on the right of the masked items. 

In a recent competition, the NVIDIA RAPIDS.AI team won the ACM WSDM 2021 WebTour Workshop Challenge, organized by Booking.com\cite{goldenberg2021booking}. The task was predicting the last booked city in a multi-destination trip and we treated it as a session-based recommendation problem. Our solution\cite{schifferer2021} for that competition used an ensemble of a Transformer (XLNET), a GRU-based and an MLP-based architectures.

Our solution for this SIGIR 2021 Workshop on E-commerce Data Challenge was purely based on Transformers, challenging once more the status-quo of recent RecSys competitions that were won by non-deep learning solutions \cite{dietmar2020why}. We employed causal LM and masked LM training approaches with Transformer-XL and XLNet, respectively, which are described next.


Transformer-XL \cite{transformerxl2019} is a \emph{causal language model} that solves context fragmentation by introducing a segment recurrence mechanism where the hidden states of previous blocks are cached and used to extend the context of the new upcoming segment. 

XLNet \cite{xlnet2019} takes advantage of both causal and masked language modeling, while addressing their limitations, by defining a new training objective called \emph{permutation Language Modeling}. In this work, we trained XLNet with masked LM, as we empirically found that it provided higher accuracy compared with its original permutation LM, maybe because of the shorter session length. 

For the competition, as the majority of sessions are short we feed the user session sequences truncated to the last 30 interactions to the Transformer models and do not use the segment recurrence mechanism defined in Transformer-XL and XLNet. 

\vspace{-3.0mm}

\section{The Challenge}
\label{sec:challenge}

The SIGIR 2021 Workshop on E-commerce Data Challenge \cite{tagliabue2021sigir} was organized by a partnership between academic researchers and Coveo company. The competition presented two tasks: (1) session-based recommendation and (2) cart-abandonment.

We competed for both tasks and open-source the code and documentation of our solutions on GitHub\footnote{\url{https://github.com/NVIDIA-Merlin/competitions/tree/main/SIGIR_eCommerce_Challenge_2021}}. For space reasons, this paper describes only our solution for task (1), but a detailed report on our approach for task (2) can be found in that GitHub repository.

For the recommendation task (1), the models were evaluated for their ability to predict the \emph{immediate next product} interacted by the user in a session (\emph{Mean Reciprocal Rank - MRR}) and to predict \emph{all subsequent interacted products} in the session, up to a maximum of 20 after the current event (\emph{F1 score}).

\subsection{Competition Dataset}

This competition provided a very rich dataset in terms of the diversity of data that could be relevant for personalized recommendation in e-commerce. It contains more than 37 million events distributed in almost 5 million sessions and is composed of three tables: (1) browsing events, (2) search events and (3) sku content. 

The browsing table (1) contains logs of user events on product pages (view, detail, add-to-card, remove-from-cart and purchases) and also views of non-product pages (page views) like FAQ and promotions. The search table (2) contains search events associated with sessions and includes a query vector generated from the text of the search terms, a list of the products presented to the user in the search results and the products that were clicked by the user. Finally, the SKU content table (3) provides product metadata information, like the quantized price, the product category and vectors representing the text and image of the product. 

In particular, as can be seen in Appendix~\ref{sec:sim_analysis}, we could observe a high average intra-session similarity of interacted product description and image vectors. Under the assumption that users browse on similar items in their session, it is clear that those pre-trained vectors are really able to capture the semantics of the products textual and image data.

\section{Pre-processing and Feature Engineering}
\label{sec:preproc_feat_eng}

In this section we describe our approach for preprocessing and feature engineering, which was implemented using RAPIDS cuDF, a library for GPU-accelerated dataframe
transformations and NVTabular, a library for tabular data preprocessing specialized for recommender systems.

\vspace{-3.0mm}

\subsection{Data Augmentation}
For the recommendation task models are evaluated by their ability to predict the next products the user would interact in the continuation of their sessions, and only product events were expected to be predicted. The percentage of page view events on non-product URLs (~70\%) was higher than product browsing events (~28\%) and search click events (~2\%) for the train set. We augmented the sequence of product events with page views and encoded page views URLs together with product SKUs in a single categorical feature, as if page view URLs were \emph{virtual} products. This was especially useful for sessions where only page view events were available. In addition we also included in the sequences the search clicks events which were available in the search table.
Using this augmentation approach allowed the models to capture more fine-grained sequential patterns and improved their recommendation accuracy.

\subsection{Feature Engineering}
In e-commerce datasets users may interact many times with the same product in different ways, e.g., by clicking, checking product details, adding and removing from the cart, or purchasing. In this dataset 13\% of interaction are repeated within sessions.

Recommendations were evaluated by the ability to predict the next item the user will interact with regardless of event type. To address this we kept only the first event with a product and summarized the level of interest of the user in such product by means of other features: number of interactions in the same product within the session, and flags about whether the user has checked product details or whether the product was added to cart within the session.

We encoded the following categorical features as contiguous ids, to be efficiently represented by embedding tables: event type, product category and sub-category, bucketed price, and item ids encoding both product SKUs and page views URLs (\emph{virtual} items). 

We also created a feature by dividing the bucketed product price by its average for the product category and sub-category, so that patterns about browsing on more expensive or cheaper products within the category can be detected. Finally, we created temporal features to encode the product recency -- elapsed time since product was first seen, $ \text{log}(\text{days}) $ --, and also encoded hour of the day and day of the week by encoded with cycling transformation, i.e., using \emph{sine} and \emph{cosine} functions. All these numerical features were normalized using standardization, except the cycling features which are already uniformly distributed between -1 and 1. Finally all interaction features were grouped by sessions, so that each training example represented one session.

\subsection{Frequency Capping of Unpopular Products}
\label{sec:freq_cap}
We analyzed the distribution of the item's frequency and observed that it follows the typical long-tail we find in e-commerce interactions datasets \footnote{For example, the Gini index of the item frequency distributions for the Coveo dataset is 0.8849 and for the YOOCHOOSE dataset\cite{ben2015recsys} is 0.8987}, where a small number of popular products concentrate most of the interactions. As an example, from the 57,483 available product SKUs, the top-1\% and top-5\% most popular products account for respectively 34\% and 67\% of the interactions. And 50\% (median) of the items receive at most 5 interactions. 

As it is hard to learn meaningful item embeddings for the unpopular items, we created a variation of the preprocessed dataset which encodes all product SKUs with less than 5 interactions into the same item id. It is important to note this approach was done after the feature engineering, so that the other features of infrequent items (e.g. category, price) remain the same. This \emph{frequency-capped dataset} was used to train one of the ensembled models.

\subsection{Data splitting for Cross-Validation}
\label{sec:data-split}

The train set covered 3 months of user interactions data and the test set covered the subsequent month.

We observed from data analysis that the distribution of  session length for the test set was close to the half of session length for the train set. Based on that, we reserved for validation the last 3 weeks of the train set. In addition, we split validation  sessions into two halves: the first half for inference and the second half for metrics evaluation.

The sessions of our train, validation and test sets were then split into 5-folds, to allow for better cross-validation and also for more diverse ensembling for allowing training models with different chunks (Out-Of-Fold) of data.

\section{Model architectures, training and evaluation}
\label{sec:model_train_eval}

\subsection{Multi-modal features processing}

The tabular features were composed by categorical embedded features and numerical features represented as continuous values. We apply \emph{layer normalization} \cite{ba2016layer} to all features individually, as we empirically observed that without it the model accuracy decreased when combining numerical features with categorical features, despite the fact that numerical features had already been standardized during preprocessing.

For the pre-trained vectors provided in the dataset based on text (search query and product description vectors) and image (product image vector) we found out that it was better to apply L2-normalization rather than using the original vectors, either with layer normalization. L2-normalization makes the feature scale similar, but also preserves the similarity relationships between similar products, aligned with \cite{moreira2019importance}.

\subsection{Model architecture}
\label{sec:architecture}

\begin{figure}[ht]
  \centering
  \includegraphics[width=0.50\textwidth]{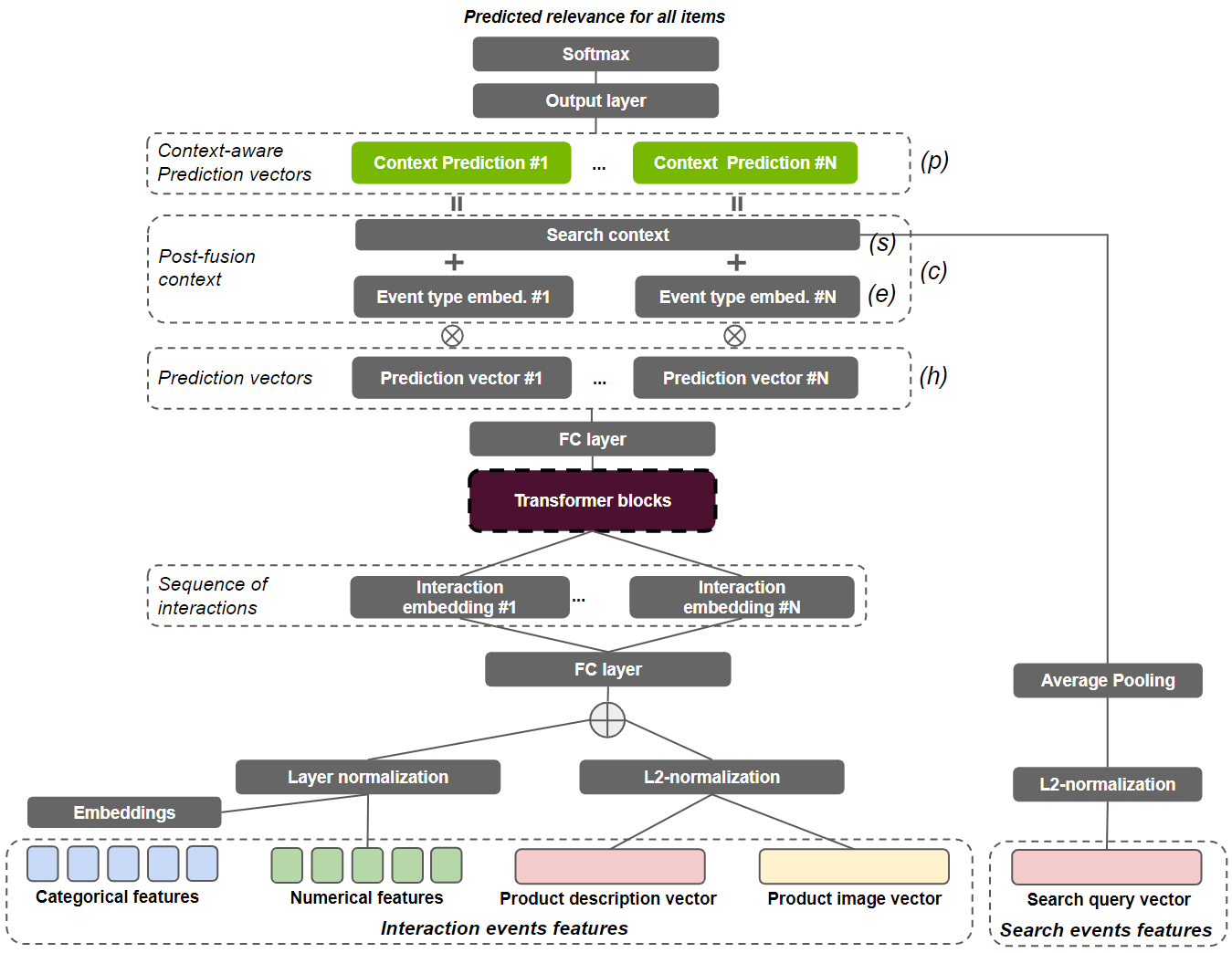}
  \caption{Our base Transformer architecture}
  \label{fig:architecture}
  
\end{figure}

Our base neural network architecture is presented in Figure~\ref{fig:architecture}. From bottom-up, all interaction features are normalized, concatenated and combined by a Fully Connected (FC) layer to produce an \emph{interaction embedding}. The sequence of \emph{interaction embedding} is fed to a Transformer architecture (Transformer-XL or XLNET), which outputs a vector for each position in the sequence. Those outputs are then projected by a FC layer to prediction vectors $ h \in  R^d $.

To provide contextual information about the item to be predicted, we leveraged the \emph{Latent Cross} \cite{beutel2018latent} technique. We create \emph{context-aware prediction vectors} $ p $ by combining the prediction vectors $ h $ with a \emph{post-fusion} contextual vector $ c $ using element-wise multiplication $ p = h \odot (1 + c) $. As the contextual vector mean is close to zero, summing 1 will center the multiplicative element distribution around 1 and can act as a mask over $ h $. The contextual vector is computed by $ c = e + s + f $. All vectors $h$, $c$, $e$, $s$, $f$ have the same dimension $d$ to allow for element-wise operations. The \emph{search context} $ s $ is calculated as the average of search query vectors (if search queries happened in the session or a zeroed vector otherwise). 

The $ f $ vector represents an embedding of a boolean feature that indicates whether the item to be predicted is a infrequent item or not. This vector is only used when training on the dataset variant with item id frequency-capping (\cref{sec:freq_cap}), to provide context on whether the item to be predicted is an infrequent item, so that the network can make the trivial prediction of the infrequent item id (i.e. 1) in such cases. During inference we set that boolean feature to \emph{frequent item}, so that the network ignores the infrequent item id.

Finally, the event type embeddings $ e $ provides context about what is the event type associated to the item to be predicted (whether it is a product, search or page view event). During inference, we always set the event type to \emph{product event}, as we are only interested in predicting items that correspond to product SKUs, and not \emph{virtual} products that are actually page views URLs. With this approach, we increased the percentage of product SKUs among the top-100 recommended items during evaluation from 35\% to 90\% \footnote{The rest of page view URLs that are eventually among the top recommended ones are removed in a post-filtering step.}.

In the output layer, we use the \emph{tying embeddings} technique originally proposed for NLP \cite{inan2016tying,press2017using}, in which we share the weights of the item id embedding table with the output layer, followed by a \emph{softmax} layer to predict the relevance scores over all items. That is possible because the context-aware prediction vectors $p$ and item id embeddings have dimension $d$. This technique was originally applied for RecSys in \cite{hidasi2018recurrent} and demonstrated to be especially effective in the NVIDIA.AI team solution for Booking.com Challenge \cite{schifferer2021}.

We treat the recommendation as a multi-class classification problem and use cross-entropy loss.


\vspace{-3.0mm}

\subsection{Training and evaluation approach}

In real machine learning projects, we cannot use future data which is not available at inference time. But for machine learning competitions all available data is generally used, including test set data.

We observed that 6.7\% of items present of the test set were not seen before in the train set. Thus, we decided to include the public part (first half) of test sessions in the training, so that we could learn embeddings for recently released products. 

\begin{figure}[h]
  \centering
  \includegraphics[width=0.40\textwidth]{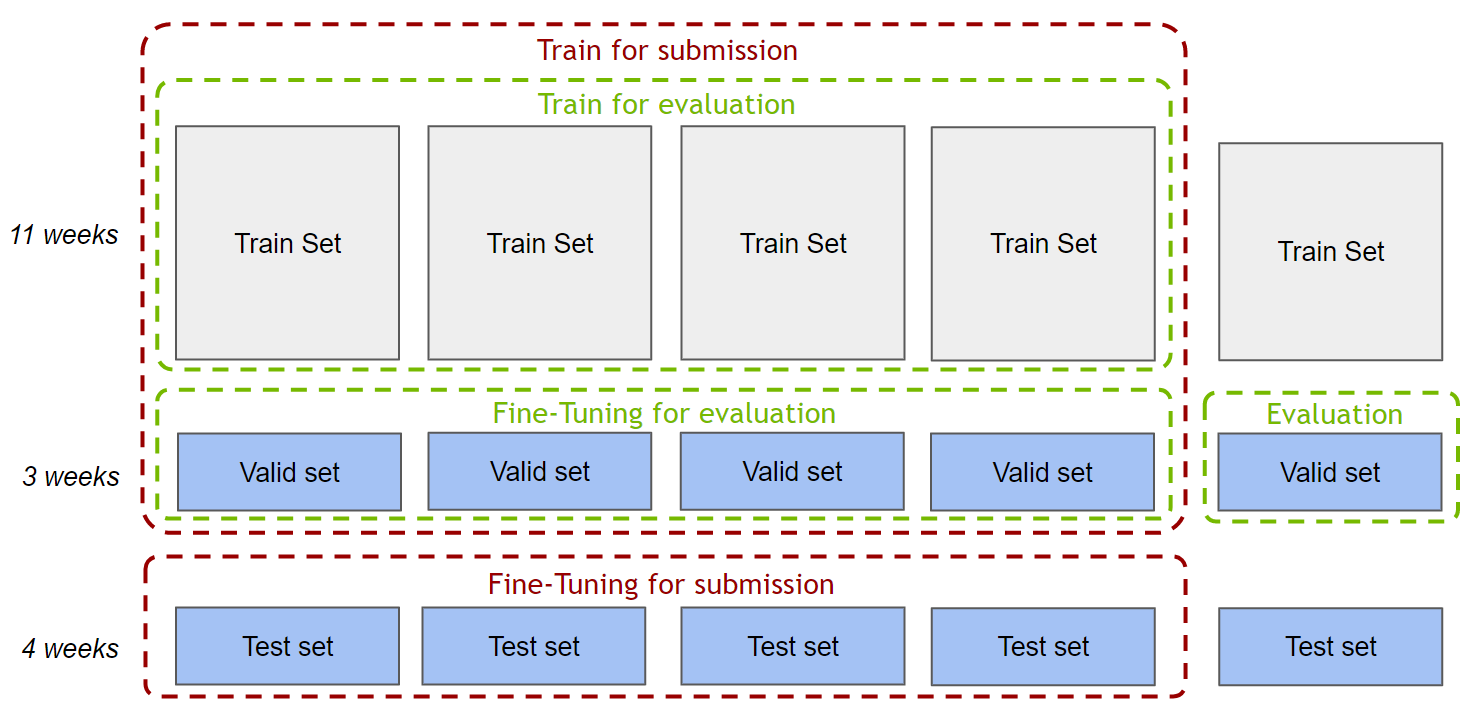}
  \caption{Illustration of training and evaluation strategy}
  \label{fig:train_eval_strategy}
\end{figure}

We tried three approaches to train on test set: (1) concatenating train, validation and test sets and shuffling, (2) same as (1), but sorting data by time, and (3) pre-training with train and validation set and fine-tuning only the item embeddings using the test set. The latter approach performed the best (2\% better than the other approaches) and is illustrated in Figure~\ref{fig:train_eval_strategy}. 

\vspace{-4.0mm}

For fine-tuning we freeze all the weights of the network except the item id embeddings, whose weights are shared with the output layer. For our evaluation setup, we perform an analogous approach, pre-training models with just training data and fine-tuning with validation data. Only the first half of validation sessions are used for fine-tuning, so that they are compatible with test sessions length.

Models are trained using a 5-fold strategy (\cref{sec:data-split}), meaning that for each validation fold a model is trained using Out-Of-Fold (OOF) sessions, corresponding to about 80\% data. 

The full pipeline (pre-training, fine-tuning, evaluation and prediction over the test set) runtime for each model had an average 265 min with 53 min std., in an instance with one V100 GPU with 32 GB of memory and 8 CPUs. In particular, the models throughput during inference for next-click prediction is of about 800 sessions / sec. with this single GPU hardware.

\section{Ensembling and Results}
\label{sec:ensembling}

We have used in our ensemble four variations of the base neural architecture presented in \cref{sec:architecture}. All neural architectures use tabular features and  product description vectors, but vary as follows:

\begin{itemize}
    \item \textbf{XLNET-IM} - XLNET with image vectors;
    \item \textbf{XLNET-S} - XLNET with search context (post-fusion);
    \item \textbf{XLNET-IM-FC} - XLNET with image vectors and item frequency capping (described in \cref{sec:freq_cap}); and
    \item \textbf{TransfoXL-IM} - Transformer-XL with image vectors.
\end{itemize}

Each of those four architectures were trained using three different hyperparameter configurations that performed well in our CV scores after hyperparameter tuning. The hyperparameters used for our models, the corresponding command lines and source code for reproducibility are available in the solution GitHub repository. 

The \emph{full ensemble} was composed of the test set predictions provided by 4 architectures x 3 hyperparameter configurations x 5 folds which results in 60 models. For each of those models, we saved the top-100 recommended items for each session and used weighted sum to produce the final recommendation lists. 

In fact, some of the predictions for our \emph{final ensemble} finished on top of the deadline hour and we had a last-minute memory issue when ensembling those large prediction files. Thus our \emph{final ensemble} submission used only models from 2 folds (24 models), which scored 1st in the Leaderboard (LB) for F1 ($0.0744$) and 2nd in the MRR ($0.2771$) LB, very close to 1st place. As soon as the LB opened again the next day for probing, we submitted our \emph{full ensemble} which scored $0.0747$ for F1 and $0.2783$ for MRR, which would have placed 1st for that metric too. 

We present in Table~\ref{tab:model_resuls}  the LB scores for each architecture / its three hyperparameters configurations (suffixed by 1,2, and 3), after ensembling their 5-folds predictions. We also present the \emph{final ensemble} and \emph{full ensemble} LB results. It can be observed that the top-4 single models were XLNET-IM-2, XLNET-IM-3, XLNET-IM-FC-2, XLNET-IM-FC-3, which all include the product image vectors as input features. Embedding the original item ids (XLNET-IM-2, XLNET-IM-3) worked better than applying frequency capping (XLNET-IM-FC-2, XLNET-IM-FC-3). It can also be observed that the Transformer-XL, trained with causal LM approach, performed worse than XLNet which was trained with masked LM. Finally, it can be seen that \emph{full ensemble} improved the best single model LB score by 1.4\% for MRR and 0.9\% for F1.

\begin{table}[ht]
\caption{LB scores for individual architectures (5-folds ensembles) and for the full ensemble}
\footnotesize
\begin{tabular}{lrr}
Model & MRR & F1 \\ \hline
XLNET-IM-1     & 0.2638               & 0.0712 \\
XLNET-IM-2     & 0.2746               & 0.0734 \\
XLNET-IM-3     & 0.2741               & 0.0741 \\ \hline
XLNET-S-1      & 0.2634               & 0.0712 \\
XLNET-S-2      & 0.2669               & 0.0713 \\
XLNET-S-3      & 0.2671               & 0.0719  \\ \hline
XLNET-IM-FC-1  & 0.2640               & 0.0714   \\
XLNET-IM-FC-2  & 0.2726               & 0.0732   \\
XLNET-IM-FC-3  & 0.2693               & 0.0724  \\ \hline
TransfoXL-IM-1 & 0.2317 & 0.0614  \\
TransfoXL-IM-2 & 0.2421 & 0.0632  \\
TransfoXL-IM-3 & 0.2406 & 0.0629 \\ \hline
\textbf{Final Ensemble (24 models)}  & \textbf{0.2771}               & \textbf{0.0744} \\ 
\textbf{Full Ensemble (60 models)}  & \textbf{0.2784}               & \textbf{0.0748} \\ \hline
\end{tabular}
\label{tab:model_resuls}

\end{table}

We present an analysis of the \emph{full ensemble} predictions to better understand the recommendation accuracy with respect to different characteristics of sessions and items in Appendix~\ref{sec:pred_analysis}.

\section{Conclusion}

In this paper, we present one of the winning solutions for the session-based recommendation task of SIGIR 2021 Workshop on E-commerce Data Challenge.
Our solution leveraged an ensemble of Transformers models, which effectively learned sequential patterns on users browsing to predict the next interacted items. The proposed architecture leveraged multi-modal information from tabular, textual and image data for more accurate recommendations. Finally, in our prediction analysis (Appendix~\ref{sec:pred_analysis}) we observed how the item popularity-bias affects the recommendation accuracy.

\appendix

\vspace{-1.0mm}

\section{Similarity analysis of product description and image vectors}
\label{sec:sim_analysis}

We performed an analysis of how similar where products interacted by the users within their browsing sessions, in terms of the average similarity of their textual descriptions and product images vectors.

\begin{figure}[ht]
  \centering
  \includegraphics[width=0.40\textwidth]{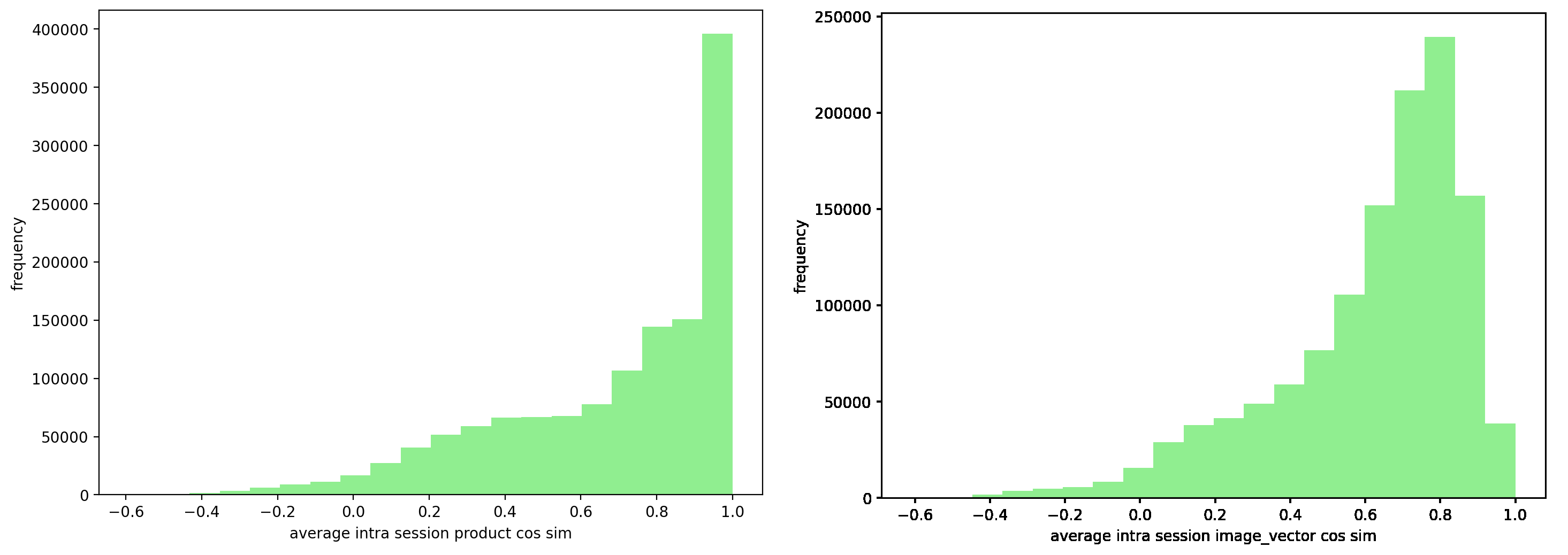}
  \caption{Distribution of avg. intra-session similarity of product description (left) and product image vectors (right)}
  \label{fig:sim_hist}
\end{figure}

It can be seen in Figure~\ref{fig:sim_hist} that the average intra-session similarity, i.e., for each pair of products within session, is generally high for both description vectors and image vectors. Under the assumption that interactions within sessions occur on similar items, it is clear that such pre-trained vectors were able to capture the semantics of textual and image data associated to the product, making them informative features to be used for next-click prediction.

\begin{figure}[ht]
  \centering
  \includegraphics[width=0.40\textwidth]{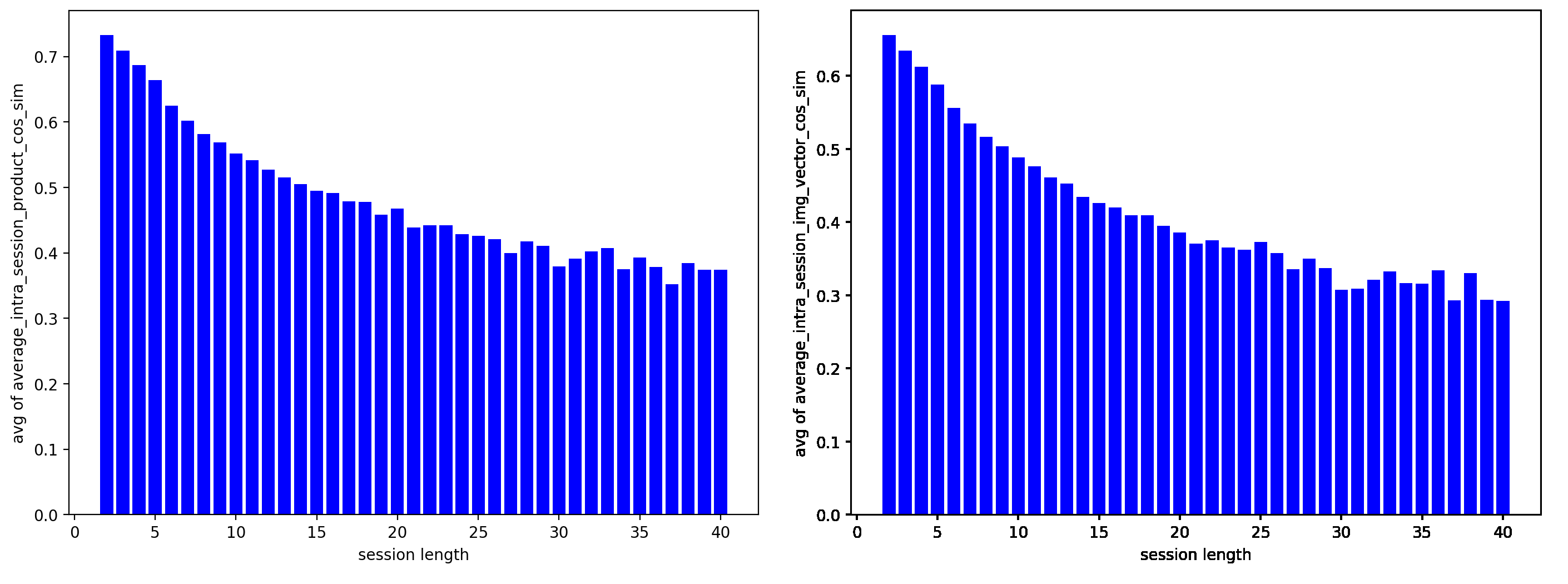}
  \caption{Avg. intra-session similarity of product description (left) and product image vectors (right) per session length}
  \label{fig:sim_by_session}
\end{figure}

In Figure~\ref{fig:sim_by_session}, it is interesting to observe how the avg. intra-session similarity based on both product descriptions and images decreases for longer session. That might indicate that the longer the sessions the less specific or targeted is the user browsing, as the interacted products are more diverse. This fact might be related to the lower recommendation accuracy we observe in longer sessions, which we discuss in Appendix~\ref{sec:pred_analysis}.

\section{Prediction analysis}
\label{sec:pred_analysis}

We present some predictions analyses that help to understand better models accuracy with respect to different characteristics of sessions and items to be predicted. So we used predictions over the validation set from the full ensemble, where the first half of each session is used for inference and the second half for computing  metrics.

\begin{figure}[ht]
  \centering
  \includegraphics[width=0.40\textwidth]{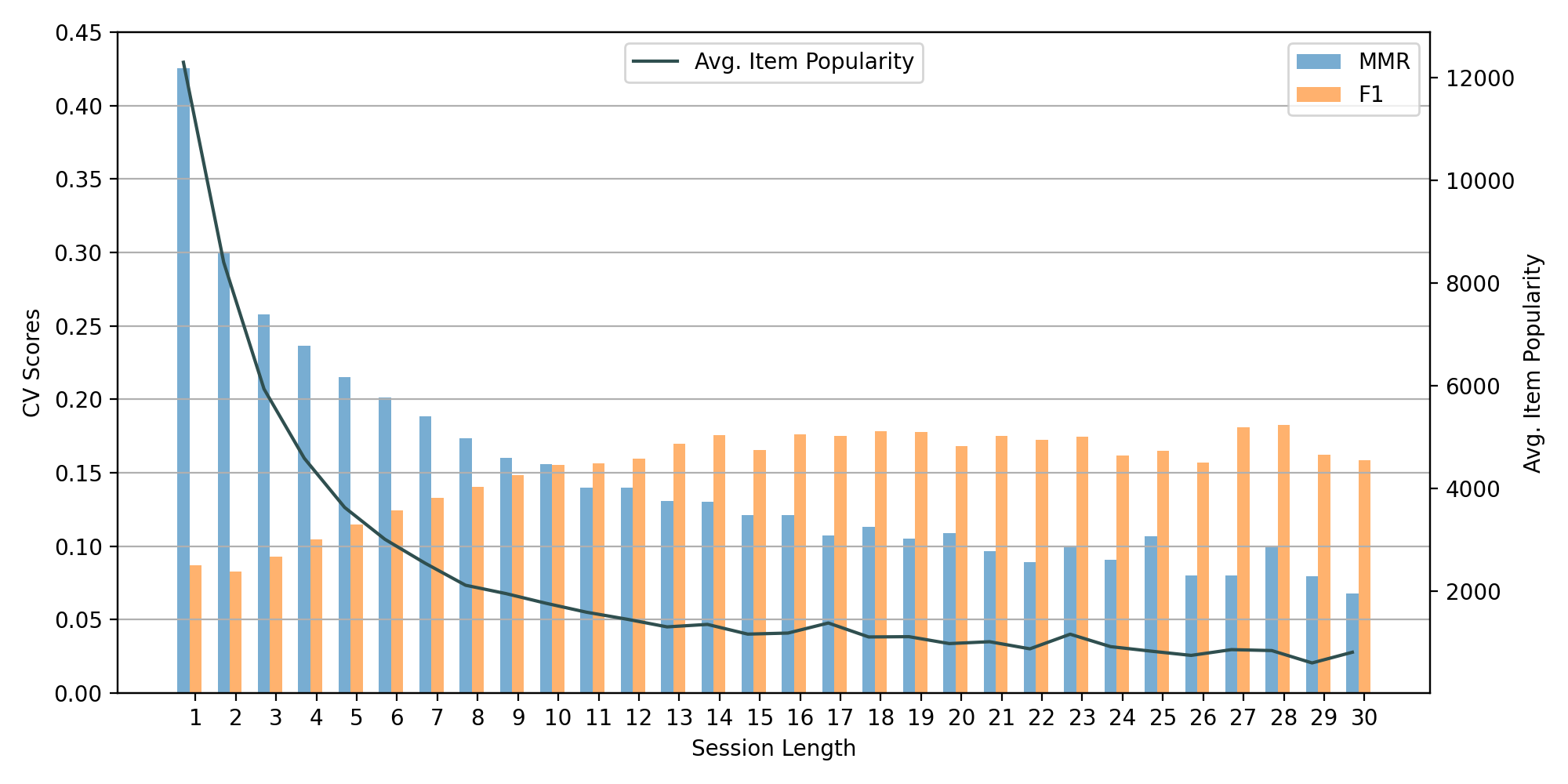}
  \caption{Accuracy (MRR and F1) x Session Length}
  \label{fig:acc_session_length}
\end{figure}

In Figure~\ref{fig:acc_session_length}, we can observe that the F1 score is in general higher for longer sessions, which makes sense, as there are increased chances of the recommended items be present in the second half of the sessions.  On the other hand, it is possible to observe that MRR, which measures the ability to predict the immediate next item, decreases for longer sessions. This might seem counterintuitive, as for longer sessions there is more contextual information available for recommendation. We can observe that popularity-bias (dark gray line) plays a role here, as first user interactions target more popular items. It might be the case that first session clicks come from popular products highlighted in the e-commerce home page, whereas later interactions might be more related to individual user interests. This phenomena has also been observed in session-based recommendation analyses for the news domain \cite{moreira2019chameleon} (Figure 6.3).

We also believe that this finding can be somewhat related to Figure~\ref{fig:sim_by_session}, where we see that in longer sessions users browsing might be less specific (lower similarity among interacted items), making it harder to predict accurately the next user move.

\begin{figure}[ht]
  \centering
  \includegraphics[width=0.45\textwidth]{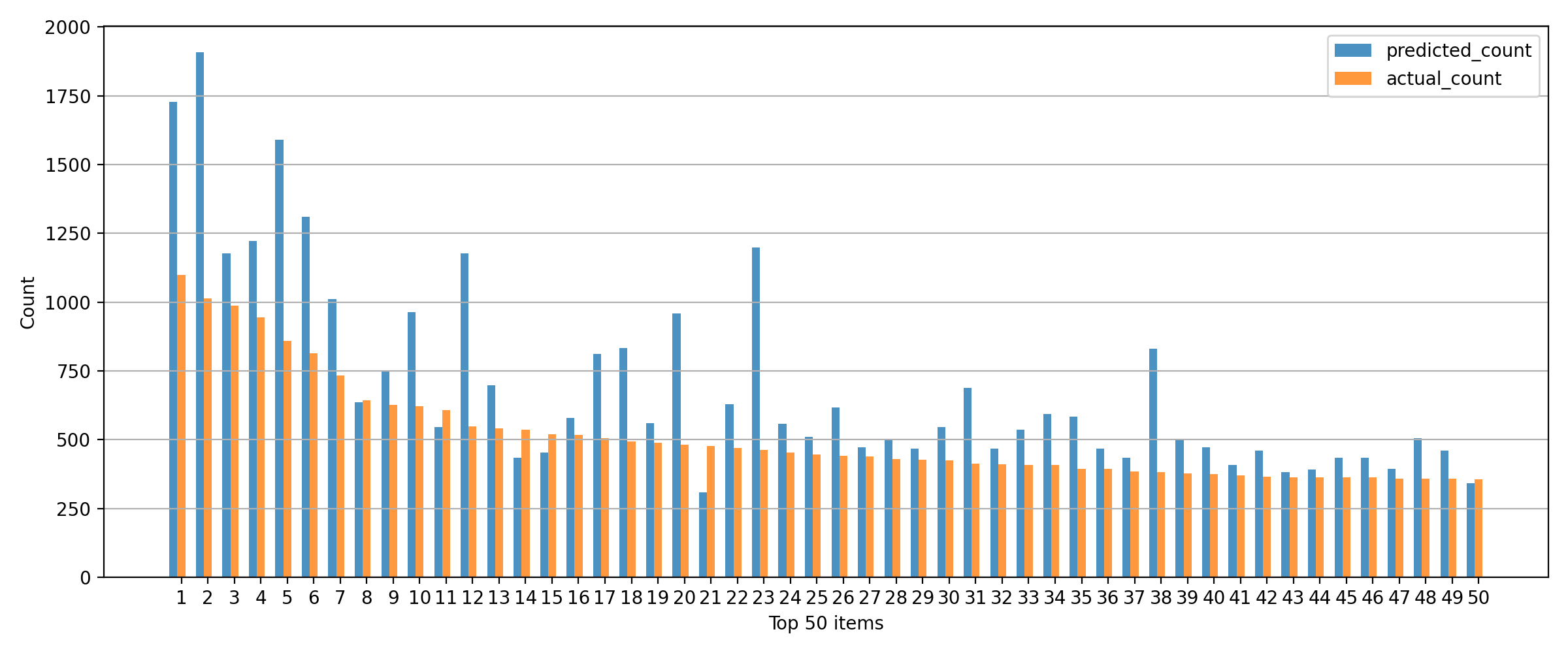}
  \caption{Frequency of popular items x frequency of top-1 predictions for such items}
  \label{fig:freq_actual_x_predicted}
\end{figure}

In Figure~\ref{fig:freq_actual_x_predicted}, we plot the frequency of the top-50 most popular products (counting only the immediate next-item of the session) in the validation set and compare with the frequency of such items as the top prediction for a session. We can notice in general a pattern of recommending popular items more often. The Pearson correlation between the actual and prediction item frequency is 0.9306 for the top-50 popular items and 0.8561 for all items.

\begin{figure}[ht]
  \centering
  \includegraphics[width=0.4\textwidth]{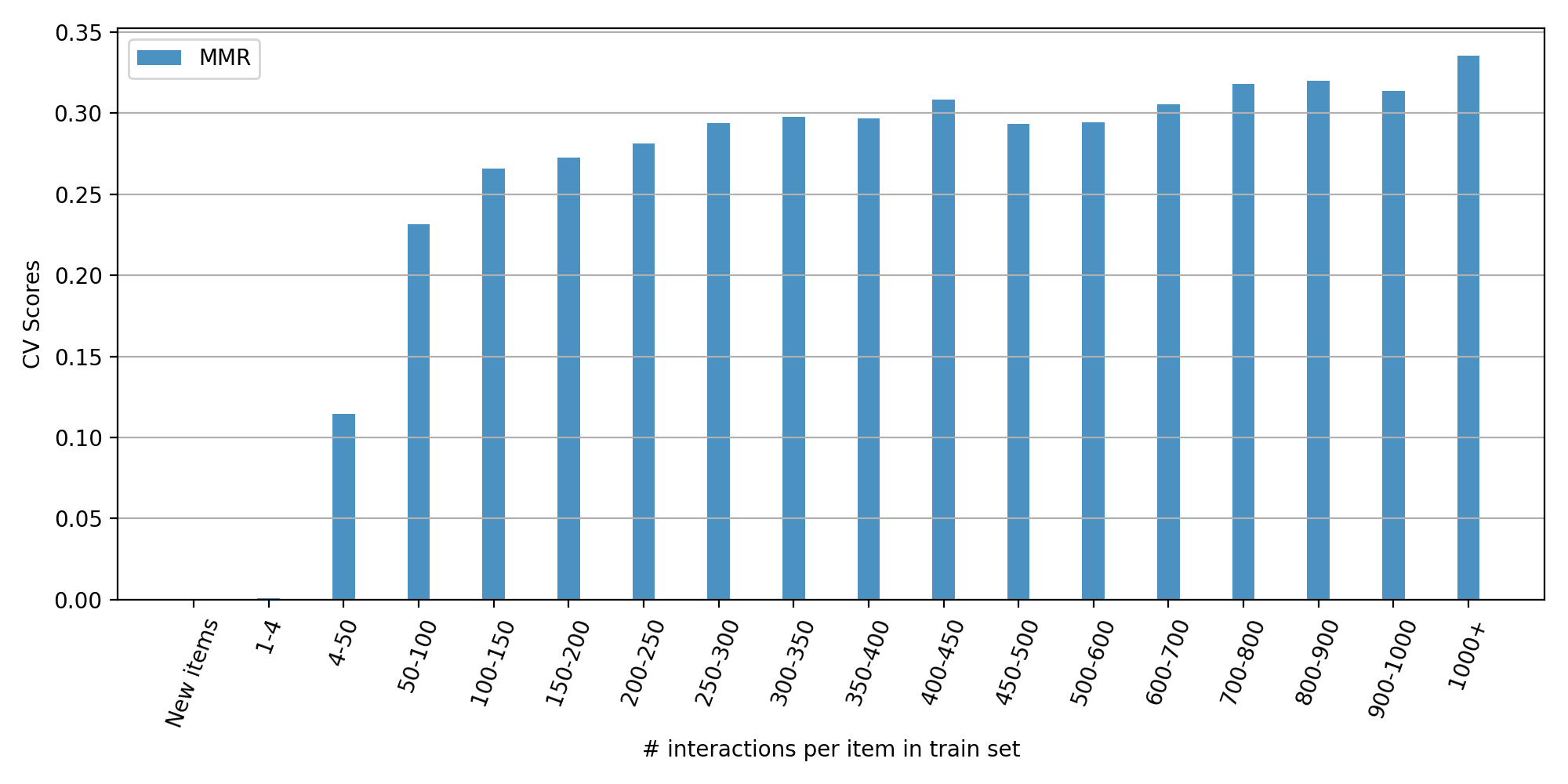}
  \caption{Accuracy (MRR) x next-item product popularity}
  \label{fig:mrr_pop}
\end{figure}

We specifically analyzed the MRR metric with respect to the popularity of the immediate next-item. It can be seen in Figure~\ref{fig:mrr_pop} that the more popular the product is, the higher its prediction accuracy. That might come from the fact that item embeddings of popular items have more chances to be trained and also because predicting very unpopular items is generally a wrong bet. Particularly, we observe a much lower accuracy (MRR=0.0009) for items whose frequency is between 1 and 4, which was one of the reasons we have included a variant of our pre-processing with item frequency capping (\cref{sec:freq_cap}). We also observe a much higher MRR when next-items are products with more than 50 interactions. Finally, "New Items" refer to products which were not seen either during pre-training or fine-tuning, as they were present only in the hidden part (second half) of validation sessions. Predictions for such unseen items got MRR=0, suggesting that alternative approaches like content-based filtering or recommending newly launched products could be considered to deal initially with the \emph{item-cold start problem}.

\begin{figure}[ht]
  \centering
  \includegraphics[width=0.40\textwidth]{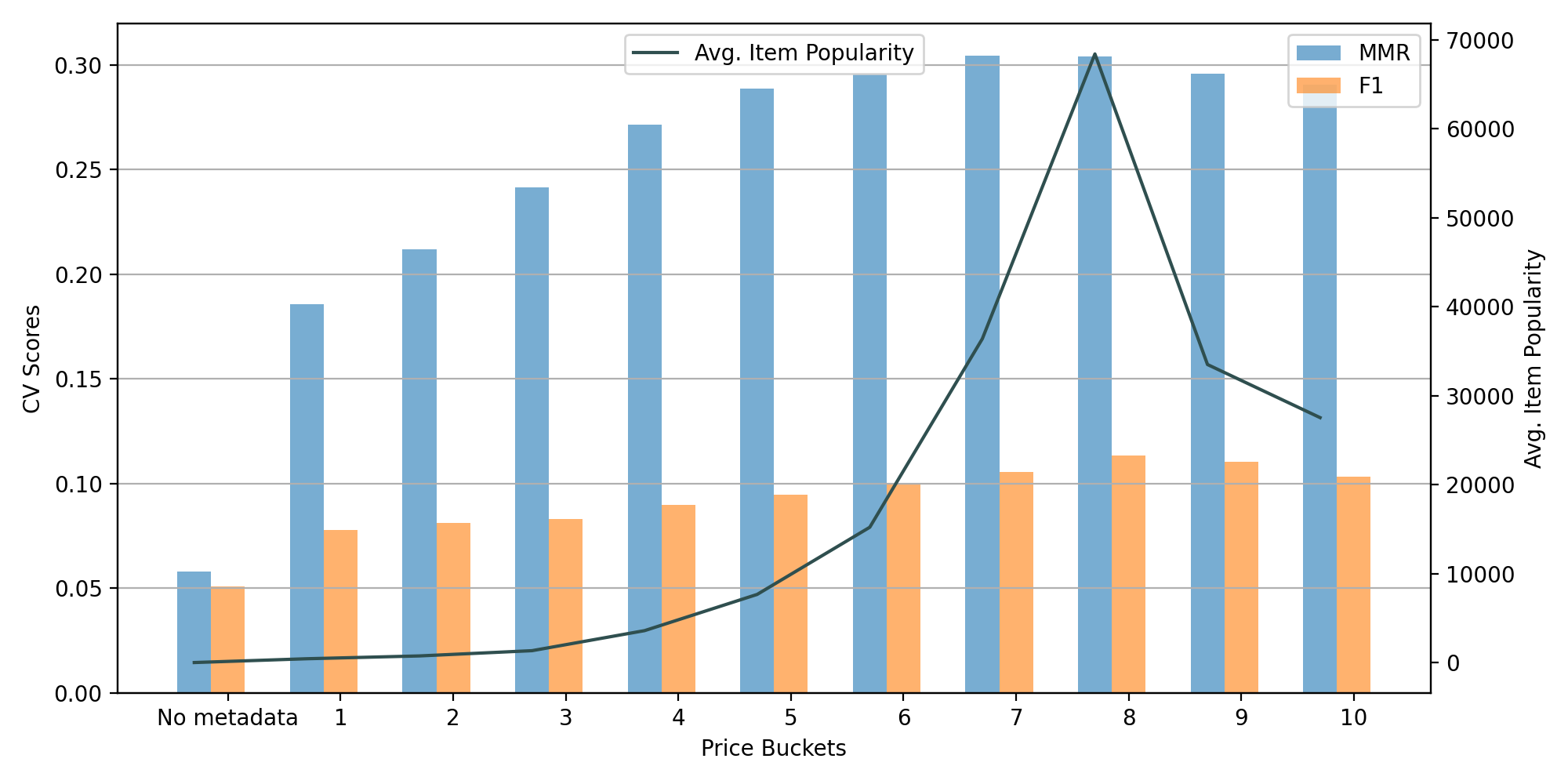}
  \caption{Accuracy (MRR) x next-item product price}
  \label{fig:mrr_price}
\end{figure}

Finally, in Figure~\ref{fig:mrr_price} we analyze the MRR according to the product price (bucketed) of the immediate next-item, where 0 is the bucket for products without metadata information (price, and description/image vectors). We observe a pattern that for lower-priced products  (buckets 1 to 4) the MRR is lower. Again, it might be an effect from popularity-bias (dark gray line), as more expensive products are more frequent in user interactions.

\bibliographystyle{ACM-Reference-Format}
\bibliography{references}

\end{document}